\documentclass[12pt,a4paper]{article}
\usepackage[latin2]{inputenc}
\usepackage{amsmath}
\usepackage{amsfonts}
\usepackage{amssymb}
\usepackage{graphicx}
\usepackage{slashed}
\begin{document}

\title{Effect of Noncommutativity of Space-time on Zitterbewegung}
\author{Ravikant Verma\footnote{Email:ravikant.verma@bose.res.in}~ and Aritra N Bose\footnote{Email:anb@bose.res.in}\\
{\it S. N. Bose National Centre for Basic Sciences,}\\{\it  JD Block, Sector III, Salt Lake, Kolkata 700106, India}}
\maketitle
\begin{abstract}
In this paper, we present the results of our investigation on the modification of Zitterbewegung due to the noncommutativity of the space-time. First, we study the effect of $\kappa$-deformation of the space-time on Zitterbewegung. For this, we start with the $\kappa$-deformed Dirac theory and using $\kappa$-deformed Dirac equation valid upto first order in deformation parameter $a$, we find the modification in the Zitterbewegung valid upto first order in the deformation parameter $a$. In the limit $a\rightarrow 0$, we get back the commutative result. Secondly, we find the modification in the Zitterbewegung due to the Magueijo-Smolin(MS) approach of doubly special relativity(DSR) and in the limit $E_p \rightarrow \infty$, we get back the result in the commutative space-time.

\end{abstract}

\section{Introduction}

Studies of various aspects of noncommutative space-times, and construction and analysis of physical models on different types of noncommutative space-times are being carried out vigorously in recent times\cite{nc11,nc21,nc22,nc31}. These recent activities were initiated after the observation that the low energy effective field theory of a specific string theory model is a field theory living on the Moyal space-time\cite{nc22}.  The Moyal space-time is defined such that the coordinates obey
\begin{equation}
[\hat{x}^\mu ,~\hat{x}^\nu ]=i\theta^{\mu\nu},\label{moyal}
\end{equation}
where $\theta^{\mu\nu}$ is a constant, antisymmetric matrix. The observation that a different limiting procedure of deriving low energy effective theory leads to another field theory model defined on the commutative space-time suggested the existence of a map that relates these two effective theories. This map, known as Seiberg-Witten map was also derived in\cite{nc22}. This map allowed the mapping of physical models defined in Moyal space-time to corresponding models in the commutative space-time. Functions defined on such a space-time can be mapped to functions defined on the commutative space-time and vice versa using Wigner map/Weyl-Moyal map. Compatibility requirement of the product of functions on noncommutative space-time under Weyl-Moyal map showed that one can use functions defined on commutative space-time but with a modified product rule called $\star$-product. Thus one can analyse physical models either by working directly in terms of functions defined on the Moyal space-time (where one treats these functions as operators since these are functions of $\hat{x}_\mu$ satisfying Eqn.(\ref{moyal})) or use functions of commutative coordinates but with the pointwise multiplication replaced by Moyal $\star$-product. Following the noncommutativity\cite{sny}, there were many developments in quantum gravity and  string theory, exploiting this notion of space-time whose coordinates do not commute. In the vicinity of the Planck length /energy scale, where quantum gravity effects are expected to be prominent, the structure of the space-time may undergo radical changes. For example, the space-time may become noncommutative in nature, where the space-time coordinates now become operator-valued satisfying the noncommutative algebra. Simplest examples include Moyal space-time, fuzzy sphere and $\kappa$-Minkowski space-time etc. Another possibility is to deform special theory of the relativity by incorporating the Planck length/Planck energy as a new constant of nature, apart from the speed of light as it is done in the doubly special theory of relativity(DSR). In this paper, we investigate how does the well known Zitterbewegung phenomenon\cite{zt3,zt1,zt2} get affected in these scenarios. The phenomenon of the Zitterbewegung came over in the study from the starting days of the relativistic quantum mechanics. Dirac and Schrodinger studied it and realised that the charged particle experiences Zitterbewegung due to the interference with particle-antiparticle pairs. Zitterbewegung is one of the very important results shedding light to trajectory of relativistic particle, which is not going to follow simple straight line as in the classical mechanics. It will get the oscillating term due to the the presence of the negative energy solution of Dirac equation. The oscillations are of the order of the Compton wavelength. Although one should note that Zitterbewegung typically occurs at a much lower energy scale than the Planck scale, but there has been a certain model, namely ADD model proposed by Arkani-Hamed $\emph{et. al.}$\cite{led1,led2} which introduces the concept of `Large Extra Dimensions' (LED) and predicts the fundamental energy scale to be much lower than the current Planck scale. Although any experimental evidence has not yet been obtained to support this model perhaps, due to the fact that ever the LHC is operated at a range which is small compared the predicted range for LED. However, the fundamental energy scale being lowered, Zitterbewegung may exhibit some signs of noncommutativity, as noncommutativity is expected to modify gravity as well as quantum theory at the fundamental energy scale. This makes a study of any possible modification of Zitterbewegung due to noncommutativity interesting.

This paper is organized as follows. In the next section, we give a brief summary of $\kappa$-deformed Dirac theory and using $\kappa$-deformed Dirac equation\cite{nhpl,w1} valid upto first order in the deformation parameter $a$, we analyse the effect of $\kappa$-deformation on the Zitterbewegung in section 3. In section 4, we summarise the Magueijo-Smolin(MS) approach of the doubly special theory of relativity, which is the modified theory of the special relativity and using this approach, one can write the DSR Dirac equation in MS base. Using this modified Dirac equation, we find the modification of the Zitterbewegung in the MS base and in section 5, we discuss about these two methods which we used in our investigation. We present our concluding remarks in section 6.

\section{$\kappa$-deformed Dirac Theory}

$\kappa$-Minkowski space-time\cite{nc61,kap2,kap3,kap5}  which emerged in the low energy limit of quantum gravity models\cite{A} is an example of noncommutative space-time whose coordinates obey Lie algebraic type of commutation relations, i.e.,
\begin{equation} 
[\hat{x}_\mu, \hat{x}_\nu]=i C_{{\mu}{\nu}{\lambda}}\hat{x}^\lambda.
\end{equation}
Here $C_{{\mu}{\nu}{\lambda}}=a_{\mu} \eta_{{\nu}{\lambda}} - a_{\nu} \eta_{{\mu}{\lambda}}$,~ $\eta_{{\mu}{\nu}}$ = diag(-1,1,1,1). Here $a_\mu$'s $(\mu=0,1,2,3)$ are real, dimensionful constants and characterize the deformation of the Minkowski space-time. For the $\kappa$-Minkowski space-time, $a_i =0$, $i=1,2,3$ and $a_0=a=\frac{1}{\kappa}$. Thus one can find the commutation relations between the coordinate of $\kappa$-Minkowski space-time as
\begin{equation}
[\hat{x}^i, \hat{x}^j]=0 , [\hat{x}^0, \hat{x}^i]=i a \hat{x}^i,~~  a=\frac{1}{\kappa}.
 \label{a1}
\end{equation}
The $\kappa$-deformed space-time also naturally associated with doubly special relativity or deformed special relativity(DSR)\cite{dsr1,dsr2}. Various aspects of physical models incorporating DSR were analysed in\cite{A,dsr4,dsr5}.

Different aspects of $\kappa$-deformed space-time and their implications on various models were studied in recent times\cite{kap2,kap3,kap5}. Construction and study of field theory models on $\kappa$-deformed space-time attracted wide attention in last couple of years\cite{a1,a2,eh1,w2,eh}. As in Moyal space-time, the statistics of particles on $\kappa$-deformed space-time is related to the symmetry of the space-time. Since this symmetry is described by a Hopf algebra, the statistics of particles in the $\kappa$-deformed space-time is naturally connected to this Hopf algebra\cite{u2,u4,u5}. In some of these studies, $\kappa$-deformed field theory models were constructed using fields which are functions of $\kappa$-deformed space-time and differential calculus on $\kappa$-deformed space-time was used in analysing these models. Some authors took the approach, developed by S. Meljanac et al, where they first mapped the $\kappa$-deformed space-time coordinates (and momenta) to that of commutative space-time such that the defining relations in Eqn.(\ref{a1}) are satisfied\cite{nc61}. Using this map, functions on $\kappa$-deformed space-time are mapped to commutative space-time. This allows one to map a theory on the $\kappa$-deformed space-time to an equivalent model in the commutative space-time\cite{nc61}. $\kappa$-deformed electrodynamics was analysed using this approach in\cite{elec} and deformed geodesic equation was derived in\cite{geod}. In \cite{nhpl}, $\kappa$-deformed Dirac equation was constructed and its non-relativistics limit was analysed. Modification to Newton's equation for a central force was studied in\cite{newton}. Certain bounds on the $\kappa$-deformation parameter were obtained in\cite{newton,di53}.

In this paper, we use this second approach of obtaining an equivalent commutative theories corresponding to $\kappa$-deformed models and analyse these models using well established methods\cite{nc61}. We now use this mapping of $\kappa$-deformed space-time coordinates to that of the commutative space-time and their derivatives obtained in\cite{nc61}. Once this map is obtained, functions on $\kappa$-deformed space-time are mapped to corresponding functions in the commutative space-time. This mapping of the $\kappa$-deformed coordinates also map the generators of the Poincare algebra, written in terms of the $\kappa$-deformed coordinates and their derivatives. The coordinates $\hat{x}_i$ and $\hat{x}_0$ are mapped to functions of $x_0,~x_i$ and $\partial_0,~\partial_i$ (which are coordinates of the commutative space-time and their derivatives) as follows\cite{nc61}

\begin{eqnarray}
\hat{x}_i=x_i \varphi(A),~~ \hat{x}_0=x_0 \psi(A) + i a x_i \partial_i \gamma(A),\label{a2}
\end{eqnarray}
where $A=-ia\partial_0$. 

The symmetry algebra of the underlying $\kappa$-space-time  is known as the undeformed $\kappa$-Poincare algebra. The generators $D_{\mu}$ and $M_{{\mu}{\nu}}$ obey 
\begin{equation}
[M_{{\mu}{\nu}} , D_{\lambda}]=\eta_{{\nu}{\lambda}} D_\mu - \eta_{{\mu}{\lambda}} D_\nu ,~~ [D_\mu , D_\nu]=0,\label{dirac4}
\end{equation}
\begin{equation}
[M_{{\mu}{\nu}}, M_{{\lambda}{\rho}}]=\eta_{{\mu}{\rho}} M_{{\nu}{\lambda}} + \eta_{{\nu}{\lambda}} M_{{\mu}{\rho}} - \eta_{{\nu}{\rho}} M_{{\mu}{\lambda}} - \eta_{{\mu}{\lambda}} M_{{\nu}{\rho}}.\label{dirac5}
\end{equation}
The explicit form of the above generators of the undeformed $\kappa$-Poincare algebra are given as
\begin{equation}
M_{ij}=x_i \partial_j - x_j\partial_i, 
\end{equation}
\begin{equation}
M_{i0}=x_i \partial_0 \varphi \frac{e^{2A}-1}{2A}-x_0 \partial_i \frac{1}{\varphi}+iax_i \nabla^2 \frac{1}{2\varphi}-iax_k \partial_k \partial_i \frac{\gamma}{\varphi}.
\end{equation}
and
\begin{equation}
D_i=\partial_i \frac{e^{-A}}{\varphi},~~ D_0=\partial_0 \frac{\sinh A}{A} + i a \nabla^2 \frac{e^{-A}}{2 \varphi^2}.\label{op} 
\end{equation}
The twisted coproduct of the generators $M_{\mu\nu}$ are given as
\begin{equation}
\bigtriangleup_\varphi (M_{ij})=M_{ij} \otimes I + I\otimes M_{ij} =\bigtriangleup_0 (M_{ij}),
\end{equation}
\begin{equation}
\bigtriangleup_\varphi (M_{i0})=M_{i0}\otimes I + e^A \otimes M_{i0} + ia\partial_j \frac{1}{\varphi(A)} \otimes M_{ij}.
\end{equation}
The algebra defined in Eqns.(\ref{dirac4},\ref{dirac5}) have the same form as the Poincare algebra in commutative space-time. But the generators of the undeformed $\kappa$-Poincare algebra are modified due to $\kappa$-deformation of the space-time. In the work reported in this section and next section, we use this undeformed $\kappa$-Poincare algebra. In the limit $a\rightarrow 0$, one get back the usual Poincare algebra.

The Casimir of this undeformed $\kappa$-Poincare algebra, $D_{\mu}D^{\mu}$ is expressed as
\begin{equation}
D_{\mu}D^{\mu}=\square (1 + \frac{a^2}{4} \square).
\end{equation}
The explicit form of the $\square$ operator is given by
\begin{equation}
\square = \nabla^2 \frac{e^{-A}}{\varphi^2} + \frac{2\partial^2_0}{A^2}(1 - \cosh A). 
\end{equation}
In terms of Dirac derivatives ($D_\mu$), $\kappa$-deformed Dirac equation, one can write
\begin{equation}
(i\gamma^\mu D_\mu + m)\Psi (x)=0,\label{ddd}
\end{equation} 
where 
\begin{equation}
\Psi (x)\sim\int d^3p ~\psi(P) e^{-ipx}.
\end{equation} 
Now Eqn.(\ref{ddd}) with the choice of $\varphi = e^{-\frac{A}{2}}$ can be rewritten upto first order in deformation parameter $a$ as
\begin{equation}
\left( i\gamma^\mu \partial_\mu - \frac{a\gamma^0}{2} \partial_i \partial_i - \frac{a\gamma^i}{2}\partial_0 \partial_i +m \right)\Psi (x) = 0.\label{dirac eqn}
\end{equation}
Also, we can write Eqn. (\ref{ddd}) in momentum space as
\begin{equation}
(\gamma^\mu P_\mu -m)\psi(P)=0,
\end{equation}
This equation has the positive and negative energy solutions as
\begin{equation}
\left. | \psi_+ (P) \right\rangle \equiv \sqrt{\frac{E^\prime +m}{2m}}\begin{pmatrix}
\tilde{\psi}_1\\
\frac{\sigma .\textbf{P}}{E^\prime +m}\tilde{\psi}_1
\end{pmatrix},~~ \left. | \psi_- (P) \right\rangle \equiv \sqrt{\frac{E^\prime +m}{2m}}\begin{pmatrix}
-\frac{\sigma .\textbf{P}}{E^\prime +m}\tilde{\psi}_2\\
\tilde{\psi}_2
\end{pmatrix},
\end{equation}
where, $E^\prime=\sqrt{|\textbf{P}|^2+m^2}$ and these two solutions satisfy
\begin{equation}
(\gamma^\mu P_\mu -m) | \psi_+ \rangle = 0,
\end{equation}
\begin{equation}
(\gamma^\mu P_\mu +m) | \psi_- \rangle = 0,
\end{equation}
and any general solution can be written as,
\begin{equation} \label{sol}
\left. | \psi \right\rangle_P = C_+ \left. | \psi_+ \right\rangle e^{-ipx} + C_- \left. | \psi_- \right\rangle e^{ipx}
\end{equation}
where $C_\pm$ are the coefficients of superposition. This is the solution of the Dirac equation which is the linear superposition of the positive and negative energy solution.
\section{Zitterbewegung in $\kappa$-deformed space-time}

In this section, we analyse the effect of $\kappa$-deformation of space-time on Zitterbewegung\cite{zt1,zt2}. For this, we start with rewriting the $\kappa$-deformed Dirac Eqn.(\ref{dirac eqn}) as
\begin{equation}
i\frac{\partial \Psi (x)}{\partial t} = \left[ \alpha^i p_i \left(1-\frac{aE}{2}\right) - \beta m -\frac{a}{2}p_i p_i\right] \Psi (x).\label{r}
\end{equation}
Here the $\kappa$-deformed free Dirac Hamiltonian is given by
\begin{equation}
H=\left[ \alpha^i p_i \left(1-\frac{aE}{2}\right) - \beta m -\frac{a}{2}p_i p_i\right], \label{H_kappa}
\end{equation}
where $\alpha^i =\gamma^0 \gamma^i$,~ $\beta=\gamma^0$ and $E=\sqrt{|\vec{p}|^2 +m^2}$. These $\alpha^i$ and $\beta$ are the matrices which satisfies the Dirac algebra as
\begin{equation}
\left\lbrace\alpha^i , \alpha^j \right\rbrace=2\delta^{ij}~~~~(i,j = 1,2,3)\nonumber
\end{equation}
\begin{equation}
\left\lbrace \alpha^i , \beta \right\rbrace=0,~~~~(\alpha^i)^2 =\beta^2=1.\label{31d}
\end{equation}
These matrices commute with momentum vector $\vec{p}$ and coordinate vector $\vec{x}$. The momentum vector $\vec{p}$ and coordinate vector $\vec{x}$ satisfies
\begin{equation}
[x^i ,x^j]=0=[p^i ,p^j],~~~~[x^i , p^j]=i\delta^{ij}.\label{i30}
\end{equation}
In the Heisenberg picture, any operator $A$ obeys the equation
\begin{equation}
\frac{dA}{dt}=i[H,A].
\end{equation}
Thus, the time dependent position operator is given by
\begin{equation}
\frac{dx^i(t)}{dt}=i[H,x^i].
\end{equation}
Now using the $\kappa$-deformed Dirac Hamiltonian in the above equation, we get
\begin{equation}
\frac{dx^i(t)}{dt}=\left(1-\frac{aE}{2}\right)\alpha^i - ap^i.\label{vel}
\end{equation}
Here it is clear that the right hand side of the above equation gives the $i^{th}$ component of the velocity operator and it is also clear that $p^i$ commutes with the Hamiltonian $H$, hence $p^i$ will not change with time. Now we define time dependent $\vec{\alpha}(t)$ and implement the Heisenberg picture
\begin{equation}
\vec{\alpha}(t)=e^{iHt}\vec{\alpha}(0) e^{-iHt},\nonumber
\end{equation}
equivalently, at the infintesimal level,
\begin{equation}
\alpha^i(\delta t)= \alpha^i + i\delta t[H,\alpha^i].\label{20}
\end{equation}
Using the $\kappa$-deformed Dirac Hamiltonian in Eqn.(\ref{20}), we get
\begin{equation}
\alpha^i(\delta t)=\alpha^i - 2i \delta t \left[i\left(1-\frac{aE}{2}\right)\sigma^{ij}p_j + m\gamma^i\right].\label{21}
\end{equation}
Then
\begin{eqnarray}
p^i-\alpha^i H &=& p^i - \alpha^i \left[ \alpha^j p_j \left(1-\frac{aE}{2}\right) - \beta m -\frac{a}{2}p_j p_j\right],\nonumber\\
&=& -i\left(1-\frac{aE}{2}\right)\sigma^{ij}p_j - m\gamma^i +\frac{aE}{2}p^i + \frac{a}{2}\alpha^i p_j p_j.\label{22}
\end{eqnarray}
Now using Eqn.(\ref{22}) in Eqn.(\ref{21}), we find
\begin{equation}
\alpha^i(\delta t)=\alpha^i + 2i \delta t \left(p^i-\alpha^i H-\frac{aE}{2}p^i - \frac{a}{2}\alpha^i p_j p_j\right),
\end{equation}
after differentiating both sides with respect to time, we get
\begin{equation}
\frac{d\alpha^i}{dt}+2i\left(H+\frac{a}{2}p_j p_j\right)\alpha^i=2i\left(p^i-\frac{aE}{2}p^i\right).
\end{equation}
By solving the above differential equation, we find
\begin{eqnarray}
\alpha^i (t) &=& p^i\left(1-\frac{aE}{2}\right)\left(H+\frac{a}{2}p_j p_j\right)^{-1}\nonumber\\
&+& \left[\alpha^i (0)-p^i\left(1-\frac{aE}{2}\right)\left(H+\frac{a}{2}p_j p_j\right)^{-1}\right]e^{-2i\left(H+\frac{a}{2}p_j p_j\right)t}.
\end{eqnarray}
Now substituting above equation in Eqn.(\ref{vel}) and then integrating with respect to time, we get
\begin{eqnarray}
x^i(t)&=& x^i(0) -ap^i t+p^i(1-aE)\left(H+\frac{a}{2}p_j p_j\right)^{-1}t\nonumber\\
&+& \frac{i}{2}\left(1-\frac{aE}{2}\right)\left[\alpha^i (0)-p^i\left(1-\frac{aE}{2}\right) \left(H+\frac{a}{2}p_j p_j\right)^{-1}\right]\nonumber\\
&\times &\left(H+\frac{a}{2}p_j p_j\right)^{-1}\left(e^{-2i\left(H+\frac{a}{2}p_j p_j\right)t}-1\right).
\end{eqnarray}
Also, we can re-write above equation as
\begin{eqnarray}
x^i(t)&=& x^i(0) -ap^i t+p^i(1-aE){H^\prime}^{-1}t\nonumber\\
&+& \frac{i}{2}\left(1-\frac{aE}{2}\right)\left[\alpha^i (0)-p^i\left(1-\frac{aE}{2}\right) {H^\prime}^{-1}\right]\nonumber\\
&\times &{H^\prime} ^{-1}\left(e^{-2iH^\prime t}-1\right),\label{fin}
\end{eqnarray}
where,  $H^\prime =\left[ \alpha^i p_i \left(1-\frac{aE}{2}\right) - \beta m \right].$ Thus from Eqn.(\ref{fin}), it is clear that $\kappa$-deformation changes the frequency of the oscillating term and upon taking the limit $a \to 0$, one gets back the familiar commutative result and and the position of the relativistic particle in the $\kappa$-deformed space-time (see Eqn.(\ref{fin})) satisfy the same Heisenberg algebra as in the commutative case given in Eqn.(\ref{i30}).

This whole analysis has been carried out at the level of operators using the Heisenberg picture. The change in the frequency of the oscillatory part that we obtain is in the operatorial sense. To relate it to any physical observations one need to take the expectation value of the $x^i(t)$ in some physical state of the particle. One can observe in the commutative case that the expectation value in positive energy eigenstate of the fermion, the non-classical term i.e. the oscillating term drops out. The oscillating term contributes only if one considers the presence of the negative energy state or a linear superposition of positive and negative energy states\cite{NE}. We follow the similar steps for our analysis.

Here we have for the positive energy states\\

$ \left\langle \psi_+ \right. | \left(\alpha^i (0)-p^i\left(1-\frac{aE}{2}\right) {H^\prime}^{-1}\right)\left. | \psi_+ \right\rangle ~~~~~~~~~~~~~~~~~~~~~~~~~~~~~~~~~~~~~~~~~~~~~$
\begin{eqnarray}
&=& \frac{E^\prime +m}{2m}\begin{pmatrix}\tilde{\psi}_1^\dagger & \tilde{\psi}_1^\dagger
\frac{\sigma .\textbf{p}(1-\frac{aE}{2})}{E^\prime +m}\end{pmatrix}\begin{pmatrix}
-\frac{\textbf{p}(1-\frac{aE}{2})}{E^\prime} & \sigma \\ \sigma & -\frac{\textbf{p}(1-\frac{aE}{2})}{E^\prime}
\end{pmatrix}\begin{pmatrix}
\tilde{\psi}_1\\
\frac{\sigma .\textbf{p}(1-\frac{aE}{2})}{E^\prime +m}\tilde{\psi}_1
\end{pmatrix}\nonumber\\
&=& \frac{E^\prime +m}{2m}\left[-\frac{\textbf{p}(1-\frac{aE}{2})}{E^\prime} \left(1+ \frac{\textbf{p}^2(1-\frac{aE}{2})^2}{(E^\prime +m)^2}\right) +\frac{2\textbf{p}(1-\frac{aE}{2})}{E^\prime +m}\right] \nonumber\\ 
&=& \frac{E^\prime +m}{2m}\left[-\frac{\textbf{p}(1-\frac{aE}{2})}{E^\prime} \frac{2E^\prime}{E^\prime +m} +\frac{2\textbf{p}(1-\frac{aE}{2})}{E^\prime +m}\right]\nonumber\\ 
&=& 0
\end{eqnarray}
where $E^{\prime 2} =|\textbf{p}|^2(1-aE) +m^2$, valid upto first order in the deformation parameter $a$. Thus, from Eqn.(\ref{fin}) it is clear that in the expectation value of the operator $x^i(t)$, oscillating term will drop out.\\
Proceeding in similar way for the negative energy state we can easily get
\begin{eqnarray}
\left\langle \psi_+ \right. | \left(\alpha^i (0)-p^i\left(1-\frac{aE}{2}\right) {H^\prime}^{-1}\right) \left. | \psi_+ \right\rangle \neq 0,
\end{eqnarray}
this shows that in the presence of the negative energy state, oscillating term will contribute.

Thus, if one considers a general solution as \eqref{sol} then it is clear that upon taking expectation value of the oscillation term in that state, only the terms involving $\left. | \psi_- \right\rangle$ will survive and if we choose $C_- = 0$, then the expectation value will become zero.

\section{Magueijo-Smolin approach of Doubly Special Relativity and Effect on Zitterbewegung}

\subsection{DSR Dirac equation in Magueijo-Smolin approach}

In this sub-section, we begin by providing a brief review of the Magueijo-Smolin approach and one obtains the DSR Dirac equation in the MS basis\cite{ms1,ms2}. The Planck length scale plays an important role in the micro-level physics like the quantum gravity, string theory and the loop quantum gravity. The special theory of relativity (STR) do not have a fundamental length scale because in STR two different observers do not measure same length at same time. But if $l_p$ is the fundamental length scale then the Planck length $l_p$ must have the same value in all inertial frames but this is contradictory to STR due to length contraction. Based on this idea, Amelino-Camelia\cite{ms3} and followed by Magueijo-Smolin\cite{ms4} developed an extended form of special theory of relativity, known as doubly special relativity. This extension requires another fundamental constant, in addition to the speed of light $c$. This second fundamental constant is expected to be related to the Planck length $l_p$ or the Planck energy scale $\kappa \propto \frac{1}{l_p}$ (or Planck energy $E_p =\sqrt{\frac{c^5 \hbar}{G}})$, which implies the microscopic structure of the space-time, called the $\kappa$-Minkowski space-time.

The general idea of quantization in commutative quantum mechanics is to introduce a correspondence principle between classical and quantum variables. This is the general strategy to make the transition from the classical theory to its quantum analogue. One can also make such a transition here for the $\kappa$-deformed Heisenberg algebra and define a deformed ``correspondence principle" for the $\kappa$-Minkowski space\cite{ms2}. Here, we use the convention $\eta_{{\mu}{\nu}}$ = diag(-1,1,1,1).

For the classical theory, the phase space variables for the Minkowski space $x^\mu$ and $p^\nu$ obeys the classical Heisenberg algebra with Poisson brackets,
\begin{equation}
\{ x^\mu, x^\nu \} = 0;~~ \{ x^\mu, p^\nu \} =  \eta^{\mu\nu};~~ \{ p^\mu, p^\nu \} = 0.
\end{equation}
The correspondence between the classical phase space variables and their quantum operator analogue is established as,
\begin{equation}
\begin{split}
\hat{x}^\mu \psi(x) & = x^\mu \psi(x)\\
\hat{p}^\mu \psi(x) & = -i\hbar \frac{\partial \psi(x)}{\partial x_\mu}\
\end{split},
\end{equation}
and these operators now obey the quantum Heisenberg algebra
\begin{equation}
[ \hat{x}^\mu, \hat{x}^\nu ] = 0;~~ [ \hat{x}^\mu, \hat{p}^\nu ] =i\hbar \eta^{\mu\nu};~~ [ \hat{p}^\mu, \hat{p}^\nu ] = 0.
\end{equation}
For the $\kappa$-deformed Minkowski space the classical Heisenberg algebra gets deformed\cite{dsr5,ms2} as,
\begin{equation}
\{ x^\mu, x^\nu \} = \lambda \left( x^\mu \eta^{0\nu} - x^\nu \eta^{0\mu} \right);~~ \{ x^\mu, p^\nu \} =  \eta^{\mu\nu};~~ \{ p^\mu, p^\nu \} = 0,
\end{equation}
where $\lambda \simeq \frac{1}{E_p}$ is the deformation parameter. It has the dimension of length and is of the order of Planck length scale. Now for this deformed classical Poisson brackets the deformed correspondence principle is defined as\cite{ms2},
\begin{equation} \label{Deformed_Corr}
\begin{split}
\hat{x}^\mu \psi(x) & = \left( x^\mu - i\hbar \lambda \eta^{0\mu} x^\nu \frac{\partial}{\partial x^\nu} \right) \psi(x)\\
\hat{p}^\mu \psi(x) & = -i\hbar \frac{\partial \psi(x)}{\partial x_\mu}\
\end{split}.
\end{equation}
And these operators then can be shown to follow the $\kappa$-deformed Heisenberg algebra for the $\kappa$-Minkowski space.
\begin{equation}
[ \hat{x}^\mu, \hat{x}^\nu ] = i\hbar \lambda \left( \hat{x}^\mu \eta^{0\nu} - \hat{x}^\nu \eta^{0\mu} \right);~~ [ \hat{x}^\mu, \hat{p}^\nu ] = -i\hbar \left( - \eta^{\mu\nu} + \lambda \hat{p}^\nu \eta^{0\mu} \right);~~ [ \hat{p}^\mu, \hat{p}^\nu ] = 0.
\end{equation}
This deformed correspondence now allows us to derive the Dirac equation from the Magueijo-Smolin dispersion relation in the same way as is taken for commutative quantum mechanics. We begin with the Maguiejo-Smolin relation\cite{ms4}
\begin{equation}
E^2 = p^2 c^2 + m^2 c^4 \left( 1 - \frac{E}{E_p} \right)^2. \label{bbb}
\end{equation}
From the deformed ``correspondence principle" \eqref{Deformed_Corr} we get the Klein-Gordon equation,
\begin{equation}
\left[ -\frac{1}{c^2} \frac{\partial^2}{\partial t^2} - \left( \frac{mc}{\hbar} (1 - \frac{i\hbar}{E_p} \frac{\partial}{\partial t}) \right)^2 \right] \phi( \vec{x} , t ) = - \Delta \phi( \vec{x} , t ).
\end{equation}
It can also be ``factorised" to derive the Dirac equation for the $\kappa$-Minkowski space in the MS base\cite{ms2},
\begin{equation}
\left[ \left( 1 + \frac{mc}{E_p} \beta \right) i \frac{\partial}{\partial t} + i \vec{\alpha}.\vec{\nabla} + \frac{mc}{\hbar} \beta \right] \Psi( \vec{x} , t ) = 0.\label{bbb1}
\end{equation}
This is the modified Dirac equation in the MS approach. In the limit $E_p \rightarrow \infty$, one gets back the result in the commutative space-time. We will use this modified Dirac equation in our analysis of the Zitterbewegung problem in the Maguiejo-Smolin energy dispersion relation base.\\
Above equation, we can write in the momentum space ($c=1=\hbar$) as
\begin{equation}
\left[\left(\gamma^0 + \frac{m}{E_p}\right)p_0 + \gamma^i p_i - m \right]\phi(p)=0,~~~p_\mu =-i\partial_\mu ,\label{kkk}
\end{equation}
and has the positive and negative energy solutions as
\begin{equation}
\left. | \phi_+ (p) \right\rangle \equiv \sqrt{\frac{\tilde{E} +m}{2m}}\begin{pmatrix}
\tilde{\phi}_1\\
\frac{\sigma .\textbf{p}}{\tilde{E} +m}\tilde{\phi}_1
\end{pmatrix},~~\left. | \phi_- (p) \right\rangle = \sqrt{\frac{\tilde{E} +m}{2m}}\begin{pmatrix}
-\frac{\sigma .\textbf{p}}{\tilde{E} +m}\tilde{\phi}_2\\
\tilde{\phi}_2
\end{pmatrix}, \label{solution}
\end{equation}
where $\tilde{E}=E+\frac{mE}{E_p}$ and then
\begin{equation}
\left. \Psi( \vec{x} , t ) \right\rangle \sim \int d^3p ~ \left( d_+ \left. | \phi_+ \right\rangle e^{-ipx}+ d_- \left. | \phi_- \right\rangle e^{ipx} \right),
\end{equation}
where $d_\pm$ are the coefficients of superposition.

\subsection{Effect on Zitterbewegung in Magueijo-Smolin basis}
In this sub-section, we find that how does MS approach of DSR effect on Zitterbewegung. Starting from the deformed dispersion relation in the MS approach of the DSR\cite{ms1,ms2} which is given as
\begin{equation} \label{EMS}
E^2 =|\textbf{p}|^2 + m^2\left( 1-\frac{E}{E_p}\right)^2,
\end{equation}
where $E_p$ is the Planck energy and  $E=\sqrt{|\vec{p}|^2 +m^2}$, is the energy of the particle.

We start with the DSR Dirac equation in the position space in MS base\cite{ms1,ms2} and corresponding deformed dispersion relation is given in the above equation. The DSR Dirac equation in MS base is given as\cite{ms2}
\begin{equation}
\left[\left(1+\frac{m\beta}{E_p}\right)i \frac{\partial}{\partial t} + i \alpha^i \partial_i +m \beta \right]\Psi =0.
\end{equation}
In the limit $E_p \rightarrow \infty$, we get back the commutative result. We can rewrite the above deformed Dirac equation as
\begin{equation}
i\frac{\partial\Psi}{\partial t}= \left[\alpha^i p_i - \beta m \left(1-\frac{E}{E_p}\right)\right]\Psi .
\end{equation}
Here, the Modified Dirac Hamiltonian in the MS base given as
\begin{equation}
H_{ms}=\alpha^i p_i - \beta m \left(1-\frac{E}{E_p}\right), \label{H_ms}
\end{equation}
where $\alpha^i$ and $\beta$ are the Dirac matrices and satisfies the Dirac algebra given in Eqn.(\ref{31d}). Here also, we proceed in the similar manner as in section 3 and with the use of the Dirac Hamiltonian in the MS basis, we get
\begin{equation}
\frac{x^i(t)}{dt}= \alpha^i.\label{47ms}
\end{equation}
Now to get the time dependent $\vec{\alpha}(t)$, we implement the Heisenberg picture as
\begin{equation}
\vec{\alpha}(t)=e^{iH_{ms}t}\vec{\alpha}(0) e^{-iH_{ms}t},\nonumber
\end{equation}
\begin{equation}
\alpha^i(\delta t)= \alpha^i + i\delta t[H_{ms},\alpha^i].
\end{equation}
Using modified Hamiltonian in the MS base in the above equation, we get
\begin{equation}
\alpha^i(\delta t)=\alpha^i - 2i \delta t \left[i\sigma^{ij}p_j + m\gamma^i\left(1-\frac{E}{E_p}\right)\right],
\end{equation}
and
\begin{equation}
p^i -\alpha^i H_{ms}=-i\sigma^{ij}p_j - m\gamma^i\left(1-\frac{E}{E_p}\right).
\end{equation}
Then
\begin{equation}
\alpha^i(\delta t)= \alpha^i + 2i\delta t(p^i - \alpha^i H_{ms}),
\end{equation}
and differentiating with respect to time, we find
\begin{equation}
\frac{d\alpha^i}{dt}+2i\alpha^i H_{ms} =2ip^i
\end{equation}
After solving the above differential equation, we get
\begin{equation}
\alpha^i(t)=p^iH^{-1}_{ms} + \left(\alpha^i (0) - p^i H^{-1}_{ms}\right) e^{-2iH_{ms}t}.
\end{equation}
Now substituting above equation in Eqn.(\ref{47ms}) and then integrating, we find
\begin{eqnarray}
x^i(t)&=&x^i(0) + p^i H^{-1}_{ms} t \nonumber\\
&+& \frac{1}{2}i \left(\alpha^i(0)
- p^iH^{-1}_{ms}\right)H^{-1}_{ms}\left(e^{-2iH_{ms}t}-1\right),\label{eee}
\end{eqnarray}
where, $H_{ms}=\alpha^i p_i - \beta m \left(1-\frac{E}{E_p}\right)$. In the limit $E_p \rightarrow \infty$, we get back the commutative result. As it is clear from Eqn.(\ref{eee}) and the form of the Hamiltonian $H_{ms}$ that the frequency of the trembling motion in this case gets modified again from the frequency in the commutative case. As explained in section 3, correspondence to any physical result should be made only by computing the expectation value of the operator $x^i (t)$. One can again show here that the oscillating term will only survive when we will take the expectation value in presence of the negative energy solutions of the Dirac equation i.e. for purely negative energy solutions or for any linear superposition of positive and negative energy solutions.\\
As in the section 3, here also proceeding in same fashion, we can easily get the expectation value of the operator $\left(\alpha^i(0)- p^iH^{-1}_{ms}\right)$ for the positive energy state (see Eqn. \eqref{solution}) to be zero because 
\begin{eqnarray}
\left\langle \phi_+ \right. | \left(\alpha^i(0)
- p^i H^{-1}_{ms}\right) | \left. \phi_+ \right\rangle = 0,
\end{eqnarray}
and in a similar way for the negative energy state we get
\begin{eqnarray}
\left\langle \phi_+ \right. | \left(\alpha^i(0)
- p^iH^{-1}_{ms}\right) | \left. \phi_+ \right\rangle \neq 0.
\end{eqnarray}
This shows that in the presence of the negative energy solution the oscillating term contributes. The position of the relativistic particle (see Eqn.(\ref{eee})) still satisfies the Heisenberg algebra same as in the commutative case given in Eqn.(\ref{i30}). The position of the relativistic particle in the $\kappa$-deformed space-time also satisfies the same algebra but the modified frequency of the trembling motion is different than the frequency obtained in MS basis, upto first order in the deformation parameter. However, this may not be true for the higher orders in the deformation parameter $a$.

\section{Discussion}
The presence of the negative energy solutions in the Dirac equation has many interesting consequences. Zitterbewegung is one of them. In this paper, we have studied how does the position of a relativistic particle get affected at time $t$ in the presence of negative energy solutions of the Dirac equation when noncommutativity is introduced. We have obtained the modifications to the Zitterbewegung(see Eqns.(\ref{fin},\ref{eee})) by considering two different models of noncommutativity. Both these models treats the effect of noncommutativity from different perspectives. They propose different energy-momentum dispersion relation and essentially describe different physics for noncommutative spaces. There are several other proposals also and one needs to compare these models in order to arrive at the correct one. We have considered two such models here, namely : (i) $\kappa$-deformed Dirac theory and (ii) DSR Dirac equation in Magueijo-Smolin approach. For both these models, we have found modifications, although the modifications are different in each model.\\

We began with the $\kappa$- Minkowski space-time. In the $\kappa$-deformed Dirac theory the dispersion relation is obtained from the Casimir by
\begin{equation}
P_\mu P^\mu=P_0 P^0 + P_i P^i =-m^2,~~~P_\mu=-iD_\mu
\end{equation}
where the explicit form of the Dirac derivative $D_\mu$ have given in Eqn.(\ref{op}). Here note that the momenta $P_0$ and $P_i$ are the $\kappa$-deformed momenta and explicitly the $\kappa$-deformed dispersion relation is given as\cite{u5}
\begin{equation}
\frac{4}{a^2}\sinh^2(\frac{ap_0}{2})-p_i^2 \frac{e^{-ap_0}}{\varphi^2(ap_0)} + \frac{a^2}{4}\left[\frac{4}{a^2}\sinh^2(\frac{ap_0}{2})-p_i^2 \frac{e^{-ap_0}}{\varphi^2(ap_0)} - \frac{a^2}{4}\right]^2=m^2.
\end{equation} 

In the limit $a\rightarrow 0$, we get the dispersion relation, $p_\mu p^\mu +m^2=0$ which is same as in the commutative space-time. The $\kappa$-deformed Dirac equation(see Eqn.(\ref{ddd})), which we used to find the correction to Zitterbewegung, satisfies the above $\kappa$-deformed dispersion relation.\\

In the second approach, we started with the Magueijo-Smolin approach \cite{ms4} and one can get the Casimir as,
\begin{equation}
m^2 = \frac{p_0 p^0 + p_i p^i}{\left( 1 - \frac{E}{E_p} \right)^2}
\end{equation}

This gives us the dispersion relation in MS base (see Eqn.\eqref{EMS}). From this dispersion relation, one can easily write the deformed Dirac equation (see Eqn.(\ref{bbb1}))\cite{ms2}. This dispersion relation follows from the undeformed homogeneous Lorentz algebra, which is similar to the $\kappa$-undeformed Lorentz algebra. But in MS base, the generators of the algebra do not get modified due to this form of deformed dispersion relation.\\

From Eqns.(\ref{fin}) and (\ref{eee}) it is easy to check that the expectation values of the operators $\left[\alpha^i (0)-p^i\left(1-\frac{aE}{2}\right) {H^\prime}^{-1}\right]$ and $\left(\alpha^i(0) - p^iH^{-1}_{ms}\right)$ are zero in the positive energy states. For positive energy solutions the expectation value satisfies classical relation, i.e. Ehrenfest's theorem holds. The non-classical contributions are non-zero only in the presence of the negative energy states. From all these, one infers that the unexpected oscillating terms occur due to the transition between positive and negative energy states.\\

There are various other approaches where the form of the homogeneous Lorentz algebra is kept same as the commutative case but the form of the generators are modified. Due to the modification in the generators, one can obtain deformed energy-momentum dispersion relation, which differs from the dispersion relation in MS base. For example, in \cite{AAA} DSR Dirac equation in the energy momentum space has been derived and it is invariant under the DSR undeformed Lorentz algebra. But the generators of the undeformed Lorentz algebra are modified due to the DSR scheme. The DSR symmetry in the $\kappa$-Minkowski space-time depends on the certain choice of the (noncommutative) differential calculus in the $\kappa$-Minkowski space-time. In later section of \cite{AAA}, a particular differential calculus was adopted and the DSR Dirac equation consistent with the $\kappa$-Minkowski space-time was derived.

\section{Conclusion}
In this paper, we presented the result of our investigations of the Zitterbewegung phenomenon from two different aspects of noncommutativity. First, we investigated that how did $\kappa$-deformation of the space-time affect the trembling motion of the relativistic particle at time t. To calculate this, we started with the $\kappa$-deformed Dirac equation valid upto first order in the deformation parameter $a$. Using the Heisenberg equation for the position operator, we calculated the position of the relativistic particle at time t which consisted of an initial position term, a term due to the motion linearly proportional to time and the unexpected fast oscillating term. This fast oscillating term is called Zitterbewegung which is now modified due to the $\kappa$-deformation of the space-time. And secondly, we presented the deformation of Zitterbewegung by using the DSR Dirac equation in the MS base. We found that the frequency of these oscillations is larger than the frequency in the commutative space-time. But for $\kappa$-deformation of the space-time, the frequency of these oscillations was less than the frequency in the commutative space-time upto first order in the deformation parameter $a$.

\section*{Acknowledgements:}
We would like to thank Prof. Biswajit Chakraborty for suggestions and discussions. Also, we would like to thank the referees for their useful comments to make this manuscript more effective.

\end{document}